\documentclass[fleqn,useAMS,usenatbib]{mnras}

\usepackage{newtxtext,newtxmath}
\usepackage[T1]{fontenc}
\usepackage[below]{placeins}
\usepackage[dvipsnames]{xcolor}

\usepackage{graphicx}	% Including figure files
\usepackage{amsmath}	% Advanced maths commands
\usepackage{chngcntr}
\usepackage{xcolor}

\newcommand{\rd}{r_{\rm cd}}
\newcommand{\rid}{r_{\rm id}}
\newcommand{\ud}{\mathrm{d}}

\graphicspath{{Figures/}}
\title[Depletion radius]{A natural boundary of dark matter haloes revealed around the minimum bias and maximum infall locations}

\author[Fong \& Han]{
Matthew Fong$^{1,2}$\thanks{E-mail: matthew.fong@sjtu.edu.cn}
and Jiaxin Han$^{1,2}$\thanks{Corresponding author: jiaxin.han@sjtu.edu.cn}
\\
$^{1}$Department of Astronomy, Shanghai Jiao Tong University, Shanghai 200240, China\\
$^{2}$Shanghai Key Laboratory for Particle Physics and Cosmology, Shanghai 200240, China\\
}

\date{Accepted XXX. Received YYY; in original form ZZZ}

\pubyear{2020}

\begin{document}
\label{firstpage}
\pagerange{\pageref{firstpage}--\pageref{lastpage}}
\maketitle

\begin{abstract}
We explore the boundary of dark matter haloes through their bias and velocity profiles. Using cosmological $N$-body simulations, we show that the bias profile exhibits a ubiquitous trough that can be interpreted as created by halo accretion that depletes material around the boundary. The inner edge of the active depletion region is marked by the location of the maximum mass inflow rate that separates a growing halo from the draining environment. This inner depletion radius can also be interpreted as the radius enclosing a highly complete population of splashback orbits, and matches the optimal exclusion radius in a halo model of the large-scale structure. The minimum of the bias trough defines a characteristic depletion radius, which is located within the infall region bounded by the inner depletion radius and the turnaround radius, while approaching the turnaround radius in low mass haloes that have stopped mass accretion. The characteristic depletion radius depends the most on halo mass and environment. It is approximately $2.5$ times the virial radius and encloses an average density of $\sim 40$ times the background density of the universe, independent on halo mass but dependent on other halo properties. The inner depletion radius is smaller by $10-20\%$ and encloses an average density of $\sim 63$ times the background density. These radii open a new window for studying the properties of haloes.
\end{abstract}


%%%%%%%%%%%%%%%%%%%%%%%%%%%%%%%%%%%%%%%%%%%%%%%%%%

%%%%%%%%%%%%%%%%% BODY OF PAPER %%%%%%%%%%%%%%%%%%

\section{Introduction}
\label{sec:introduction}
In our current understanding of structure formation, dark matter haloes are the building blocks of the large-scale structure in the Universe.
The properties and evolution of dark matter haloes are fundamental to models that describe many aspects of the Universe ranging from the galaxy formation and evolution to the overall make-up and history of the Universe. In this halo model framework, the large-scale structure can be decomposed into the distribution of haloes on large scales, convolved by the internal structure of haloes on small scales~\citep[e.g.,][]{Scoccimarro2001,Cooray2002}. Galaxies form and evolve in the potential well provided by dark matter haloes, with many intrinsic galaxy properties such as the colour, morphology and mass largely determined by the structure and evolution of the haloes they reside in~\citep[e.g.][]{Baugh2006,Benson2010,Somerville15}.

Despite the substantial work done on understanding the Universe using haloes as building blocks, practical characterisation of the fundamental properties of dark matter haloes is still subject to improvement. Most importantly, our understanding of what is the size of a halo is at best at a premature stage. Almost all of the studies about dark matter haloes so far are based on the classical virial definition of halo size derived from the spherical collapse model~\citep{Gunn1972}, which is expected to be only an approximate description of these objects. In the spherical collapse model, haloes are modelled from the collapse of a spherical initial overdensity embedded in an otherwise uniform background Universe, and the final size of the halo is defined through virialization argument. However, in the real Universe haloes are live objects that do not necessarily have a clear separation from the background Universe and are constantly accreting from and interacting with the neighbouring non-uniform environment. Furthermore, the top-hat spherical collapse model assumes that collapsing overdensities are isolated, spherical, and with no initial velocity dispersion until relaxation. This leads to the model failing to predict the evolution of simulated haloes after their turn-around epochs \citep{Suto2016}.
Such complications make the virial radius more of a practical working definition, with many variants of it defined through different overdensity criteria, such as 200 times the mean density or 200 times the critical density. Despite their popularity, such definitions do not always correctly separate the virialized part of haloes~\citep[e.g.][]{Zemp14} and can lead to apparent evolution of halo properties in absence of physical evolution~\citep{Diemer13}. 

The recently proposed splashback radius marks a significant improvement over the classical model by incorporating halo accretion into the definition of a dynamical halo boundary~\citep{Diemer2014, Adhikari2014, More2015, Shi2016} and hence has attracted substantial attentions~\citep{Snaith2017, Mansfield2017, Umetsu2017, Baxter2017, Fong2018, Chang2018,Xhakaj2020,Sugiura2020, Aung2020,Murata2020,Deason2020_galaxyEdge,Deason2020}. Another classical boundary of physical importance is the turnaround radius. The turnaround radius also arises from the spherical collapse picture, but has been relatively less studied. We summarize these boundaries below:

\begin{itemize}
    \item The virial radius, $r_{\rm vir}$, is the expected radius of a virialized halo according to the spherical collapse model. Normally this is defined through the expected virialization density, which we take from the prediction of~\citep{Bryan1998} assuming a tophat initial density in a spherical collapse model.
    
    \item The splashback radius, $r_{\rm sp}$, is practically determined to be where the density profile reaches its steepest slope. The steepening in the slope has been attributed to the build-up of particles during their first orbital apogees, where the particles have low radial velocities~\citep{Fillmore84,Bertschinger85}. This radius has gathered significant interest since its discovery as it has been shown to probe the mass accretion rate of haloes~\citep{Diemer2014, Adhikari2014, More2015, Shi2016}. Note the splashback radii of individual halo particles can span a wide range, and hence there is a large freedom in defining the overall splashback radius of a halo from particle dynamics~\citep{Mansfield2017,Diemer2017A, Diemer2017B}. In this work, we will refer to the splashback radius as the one estimated from the steepest slope location, unless explicitly specified otherwise.
    
    \item The turnaround radius, $r_{\rm ta}$, in the spherical collapse model is located where a mass shell of a halo reaches zero radial velocity before collapsing back towards the halo at a given time~\citep{Mo2010,Pavlidou2014,Tanoglidis2015,Tanoglidis2016}. For individual haloes the radial velocity profiles of particles are a combination of the peculiar and Hubble flow velocities, and the turnaround radius is located where the Hubble flow overcomes the peculiar velocity. This can potentially be used as a cosmological probe, as it reflects the competition between dark energy and gravity~\citep{Taruya2000, Falco2014, Faraoni2015, Lee2017_rta, Korkidis2019}.
\end{itemize}

Typically $r_{\rm vir} \lesssim r_{\rm sp} < r_{\rm ta}$. 
Most studies of haloes use mass or radius definitions that are within or roughly around $r_{\rm vir}$. However, the influence of haloes can go well beyond the virial radius. For example massive clusters are found to impact on the star formation rates of galaxies that are well outside their host halo's virial radius, after the galaxies have passed within the host halo's virial radius~\citep{Wetzel2014,vonDerLinden2010,Wetzel2012,Kukstas2020,Adhikari2020}. In~\citet{Bahe2013,Wetzel2014} the authors show that a significant fraction of these galaxies are not infalling for the first time and remain bound to their host haloes~\citep[see also][]{Ludlow2009}. 
Even though one can always work with a given definition of halo size, it is not difficult to imagine that simple intrinsic physical relations may become complicated or obscured in absence of a correct physical description of haloes. 

The different halo radii definitions also reflect our understanding of the halo structure in different dimensions. While the virial radius is defined through the average density expected from virialization, the splashback radius is defined through the slope of the density profile or more physically the population property of evolving orbits in an accreting halo~\citep{Diemer2017A, Diemer2017B}. 
The turnaround radius, on the other hand, can be practically defined through the velocity profile around the halo. 

In this work, we introduce two other dimensions to define a natural boundary of haloes in aim of providing a physical characterisation of haloes. We do this by first studying the bias profile, that is, the overdensity profile around a halo relative to the average overdensity profile around a random matter particle. We make use of a high resolution $N$-body simulation to extract bias profiles for haloes of different properties. The bias profile shows a typical trough around the halo boundary. This identifies the scale where the correlation between matter and haloes is the weakest, relative to the average clustering of matter, thus providing a natural dimension to define the boundary of a halo. Importantly, this bias trough is found to be close to the region of maximum mass inflow rate around the halo, signalling ongoing depletion around the bottom of the trough. These two characteristics combined lead to the interpretation of the bias trough as the location where matter is being depleted from the environment by the growing haloes over time. Correspondingly, we propose two characterisations of the trough scale: an inner depletion radius defined at the location of the maximum mass inflow rate, and a characteristic depletion radius at the minimum of the bias.

As a first paper in studying the depletion boundary, in this work we aim at establishing the concept and general characterisations of the depletion region, with some additional efforts on characterising the radius at the bias minimum. To this end we examine how the boundary features relate to the virial, splashback, and turnaround radii. By splitting haloes into bins of different halo properties, we also study how the characteristic depletion radius depends on multiple halo properties, and show that it has some simple properties such as a nearly constant enclosed density. In terms of particle orbits, we demonstrate that the inner depletion radius can be interpreted as the outermost splashback radius visually identifiable in the phase space structure around haloes. 
An immediate application of the inner depletion radius is to improve over the halo model description of the matter clustering around the quasi-linear scale~\citep[e.g.][]{Tinker2005,Hayashi08,vdB2013}, which we briefly discuss in this work. In a follow-up work, we will study the depletion process in the spatial and velocity distributions explicitly by tracking haloes over time.

This paper is organized as follows. In Section~\ref{sec:simulation} we describe the halo sample used in this work. In Section~\ref{sec:binnedBias} we introduce the characteristic depletion radius through the bias profile, examine its dependence on halo properties, and relate it with the splashback radius. In Section~\ref{sec:Rd_velocityComparison} we study the dynamical interpretation of the depletion trough and introduce the inner depletion radius through the velocity profile and the phase space structure around haloes, while relating the two radii with the turnaround radius. In Section~\ref{sec:M-r} we compare these radii and their enclosed densities to other halo boundary characterisations. In Section~\ref{sec:haloModel} we briefly discuss the implications of the halo depletion radii in the halo model. In Section~\ref{sec:conclusions} we summarize and give our conclusions.

Note that $\log$ used in this work is $\log_{10}$; unless otherwise stated all $M_{X}$ and $r_{X}$ units are in ${\rm M_{\odot}}/h$ and ${\rm Mpc}/h$, respectively; the haloes in our work are located at $z=0$.

%%%%% Introduce simulation and data used in this work
\section{Simulation and the halo sample} 
\label{sec:simulation} 
In this section we introduce the simulation and the halo sample used in this paper. Our halo and clustering data are extracted from one of the CosmicGrowth Simulations~\citep{Jing2019}, a grid of high resolution $N$-body simulations run in different cosmologies using a P$^3$M code~\citep{Jing2002}. We use the $\Lambda$CDM simulation with cosmological parameters $\Omega_{\rm m}=0.268$, $\Omega_{\Lambda}=0.732$, and $\sigma_{8}=0.831$. The box size is $600~{\rm Mpc}/h$ with $3072^{3}$ dark matter particles and softening length $\eta=0.01~{\rm Mpc}/h$. 
Groups are identified with the Friends-of-Friends (FoF) algorithm with a linking length $0.2$ times the mean particle separation. The haloes are then processed with \textsc{HBT+}\footnote{\url{https://github.com/Kambrian/HBTplus}}~\citep{Han2012,Han2018} to obtain subhaloes and their evolution histories. The resulting halo catalog has about $2\times10^{6}$ distinct haloes with masses $10^{11.5}< M_{\rm vir} [{\rm M_{\odot}}/h] < 3 \times 10^{15}$, where the minimum mass corresponds to roughly $500$ particles within the virial radius to keep a reliable resolution on the structure of haloes.

The halo sample and the list of halo properties are the same as those used in~\citet{Han2019}, which we briefly describe below. The halo centres are located at the most bound particle of the central subhalo and all haloes are at $z=0$.
\begin{itemize}
	\item $M_{\rm vir}$: The virial mass of the halo, where $M_{\rm vir}$ is the mass within a spherical volume of radius $r_{\rm vir}$ that encloses the mean density $\Delta_{\rm c}$ times the critical density of the Universe, or $M_{\rm vir} = 4\pi r_{\rm vir}^{3} \Delta_{\rm c}\rho_{\rm crit}/3$. The virial overdensity in units of the critical density, $\Delta_{\rm c}$, is predicted from the spherical collapse model~\citep{Bryan1998}. Note the total mass distribution is used in the computation of the virial mass and radius, not just the bound mass. 
	\item $V_{\rm max}$: The maximum of the circular velocity function, $V_{\rm circ}=\sqrt{GM(<r)/r}$, of the central subhalo. Because $V_{\rm max}$ is tightly correlated with halo mass, for this work we will use $V_{\rm max}/V_{\rm vir}$ to factor out the mass dependency, where $V_{\rm vir}=\sqrt{GM_{\rm vir}/r_{\rm vir}}$, and $G$ is the gravitational constant. This has also been used as a proxy for halo concentration~\citep[e.g.][]{Gao2007,Angulo2008, Sunayama2016}, and is also a description of the shape of the density profile, where $V_{\rm max}/V_{\rm vir}=1$ correspond to isothermal haloes with flat rotation curves while $V_{\rm max}/V_{\rm vir}>1$ correspond to haloes with steeper outer profiles.
	\item $j$: The spin of the central subhalo: $j=\frac{L\sqrt{|E|}}{GM^{5/2}}$~\citep{Peebles1969}, where $L$, $E$, and $M$ are the total angular momentum, energy, and mass of the central subhalo.
	\item $e$: The shape parameter of the halo, $e=e_{1}$ in this work, where $e_{i}=\lambda_{i}/\sum_{j=1}^{3}\lambda_{j}$ and $e_{1} \geq e_{2} \geq e_{3}$. The subscripts $i=1,2,3$ correspond to the three eigenvalues, $\lambda_{i}$, of the inertial tensor, the square of the three principle axes lengths of the halo mass distribution. The inertial tensor is defined as $I_{w,ij}=\sum_{p} m_{p}x_{p,i}x_{p,j}/r_{p}^{2}$, where $m_{p}$ is the mass of particle $p$, and $\vec{x_{p}}$ and $r_{p}$ are the coordinate and distance to particle $p$ relative to the halo centre. 
	\item $a_{1/2}$: The scale factor of the Universe when the halo mass was half of its final mass $M_{0}$ at $z=0$. The masses used here are calculated using the bound particles to avoid complications due to ejected or fly-by haloes. 
	\item $\delta_{\rm e}$: The halo environment defined as the matter overdensity at a halo-centric distance $r_{\rm e}\approx 1-2 {\rm Mpc}/h$ around each halo, $\delta_{\rm e}=\rho(r_{\rm e})/\rho_{\rm m}-1$~\citep[see][for more details]{Han2019}.
\end{itemize}

%%%%%%%%%% Introduce bias binned by various halo parameters
%%%%% Bias binned by halo parameters 
\section{The halo bias profile and a characteristic halo boundary} 
\label{sec:binnedBias} 
The most straightforward way to study the boundary of a halo is perhaps to examine the density profile around haloes. However, as the average density profile decreases around a halo even out to the largest scale, it is not obvious how a boundary could be defined from the density profile alone. So instead of studying the density profile itself, in this study we work with the normalized density profile of a halo, or more specifically the bias profile.

The bias profile of haloes is defined as the ratio between the halo-matter correlation function and the matter-matter correlation function. In the limit of a single halo, the halo-matter correlation function reduces to the overdensity profile of matter around the halo center. A bias profile can then be similarly defined taking the ratio between the overdensity profile and the matter-matter correlation function. Because the matter-matter correlation function is the average overdensity profile around a random matter particle, the resulting bias profile is a measure of the relative clustering around a halo compared to the clustering around a random matter particle. According to linear theory, the clustering around haloes follows the average matter clustering on large scales up to a constant bias factor. The bias profile thus factors out the background clustering profile of the Universe and leaves only the profile relevant for the haloes.

In this work we first extract individual bias profiles around each halo from the overdensity profile and the matter-matter correlation function. In principle we can proceed to analyze each bias profile individually. However, to suppress noises associated with individual profiles, we will instead focus on analyzing the stacked profiles of haloes binned in various halo properties,
\begin{equation}
	b(r) = \frac{ \xi_{\rm hm}(r) }{ \xi_{\rm mm}(r) } = \frac{\langle \delta(r) \rangle}{ \xi_{\rm mm}(r) },
	\label{eq:binnedBias}
\end{equation}
where $\xi_{\rm hm}$ and $\xi_{\rm mm}$ are the halo-matter and matter-matter correlation functions, $\delta(r) = \rho(r)/\rho_{\rm m} - 1$ is the overdensity profile of matter around each halo, $\rho_{\rm m}$ is the mean matter density of the Universe, and the averaging is over all the haloes in each halo property bin.

\begin{figure*}
	\includegraphics[width=\textwidth]{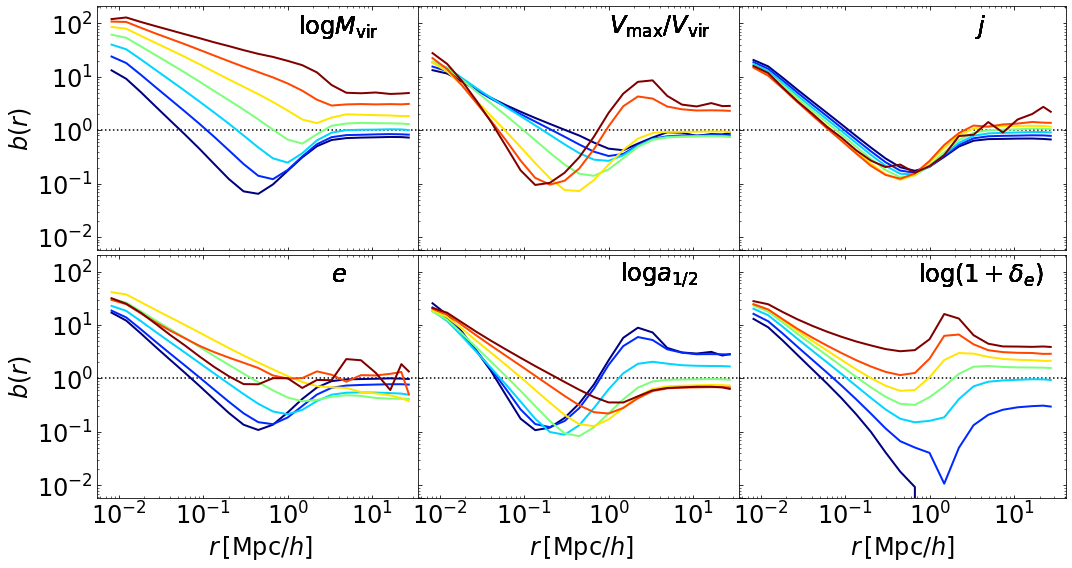}
	\caption{The halo bias profile as a function of radius, binned by various halo parameters. Each panel shows the bias binned in a different halo parameter, labelled on the top right. Each coloured line represents one binned profile. 
	In between the large-scale linear bias and the small-scale one-halo profile, there is a ubiquitous trough in nearly all the bias profiles, which defines the depletion boundary in this work.}
	\label{fig:ALL_bias}
\end{figure*}
In Figure~\ref{fig:ALL_bias} we show the bias binned by the halo parameters explored in this paper, where each panel is binned by the halo parameter labelled on the top right of each panel. 
Though the forms of the bias can be complex, there are some common features. On large scales the bias flattens to a constant value, where we expect that the halo-matter cross-correlation, or the ensemble average of the overdensities around haloes, follows the shape of the matter-matter correlation to produce the so called linear bias. In contrast to the decreasing density profile, the constant bias profile on large scales thus provides a natural reference of the background density distribution around haloes. On small scales the bias is large and dominated by the density profile of the central halo. The intermediate scale marks the transition between the one-halo feature and the large-scale background, providing the natural scale to look for the boundary of haloes. Indeed, in nearly all cases there is a clear trough in the bias in the intermediate scale, representing the location where the correlation between haloes and matter is the weakest, with respect to the correlation of matter around a random particle in the density field. The location of this bias minimum thus defines a characteristic radius of the halo, which we will call the characteristic depletion radius, $\rd$, for reasons that will become clear when discussing dynamics in Section~\ref{sec:Rd_velocityComparison}.

Beyond the bias minimum, the shape of the bias profile closely resembles the density profile of a void~\citep{Hamaus2014}, which rises with radius until it reaches a maximum in many cases. One general interpretation for this is that the mass accretion of the central halo depletes matter around it, creating a characteristic trough. Beyond the trough the influences of neighbouring haloes and the background expansion become increasingly important, and the matter distribution can also be dominated by matter in neighbouring haloes, creating the wall region that resembles the wall around voids for the same reason. In this sense, the trough can be regarded as a ``relative void" formed by the accretion of the central halo, alongside the competing mass accretion from neighbouring haloes. We will come back to this discussion in later sections. 

In the language of a halo model, the profile within the depletion boundary is thus the one-halo term, while that outside the boundary is dominated by the two-halo term. The depletion trough is thus a direct manifestation of halo exclusion. This is also supported by the location and depth of the bias trough in different bins. Taking the mass dependence as an example, the highest mass haloes can only be surrounded by lower mass neighbours. Because it is easy to tightly pack small haloes around big ones, the boundaries of the largest haloes are thus smoothly connected to the neighbourhood, with no obvious troughs created by halo exclusion. By contrast, the smallest haloes typically only find neighbours relatively farther away on halo virial scales, reflecting the difficulty in packing larger haloes around them due to halo exclusion. As we will show later in section~\ref{sec:haloModel}, the inner depletion radius that characterises the inner edge of ongoing depletion well matches the optimal halo exclusion radius in a halo model.

In order to accurately estimate the characteristic depletion radius defined in the bias profile, we first fit each bias profile with the following function,
\begin{equation}
    b^{\rm{Fit}}(r)=\frac{  1+\left(\frac{r}{r_{0}}\right)^{-(  \alpha+\beta  )}  }{  1+\left(\frac{r}{r_{1}}\right)^{-(  \beta+\gamma)  }  } \times \left(  b_{0}+\left(\frac{r}{r_{2}}\right)^{-\gamma}  \right).
	\label{eq:bFit}
\end{equation}
with $r_0<r_1<r_2$. This function has four components describing the inner-most one-halo profile ($b\propto r^{-\alpha}$ for $r \ll r_0$) before the trough, the rise beyond the trough ($b\sim r^\beta$ for $r_0<r<r_1$), the decrease after the wall ($b\sim r^{-\gamma}$ for $r>r_1$) and the large-scale linear bias ($b\sim b_0$ for $r\gg r_2$). Note the inner-most one-halo profile is better described by a NFW-like~\citep{NFW97} double power-law profile, but as we focus on studying the intermediate to large-scale feature, we will not go to that complexity and only adopt an asymptotic single power-law for the one-halo term. We find this fitting function is universal for haloes binned in different or even multiple properties, with the exception of the lowest environmental overdensity bins.% (see Figure~\ref{fig:delta_e}).} 

With this parametrization of the bias profile, in the following sections we move on with more quantitative analysis of the depletion radius, relating it to various halo properties as well as to other halo radii.

%%%%% 1D binning results
\subsection{The characteristic depletion radius}
\label{sec:haloBiasRadius}

\begin{figure} 
	\includegraphics[width=\columnwidth]{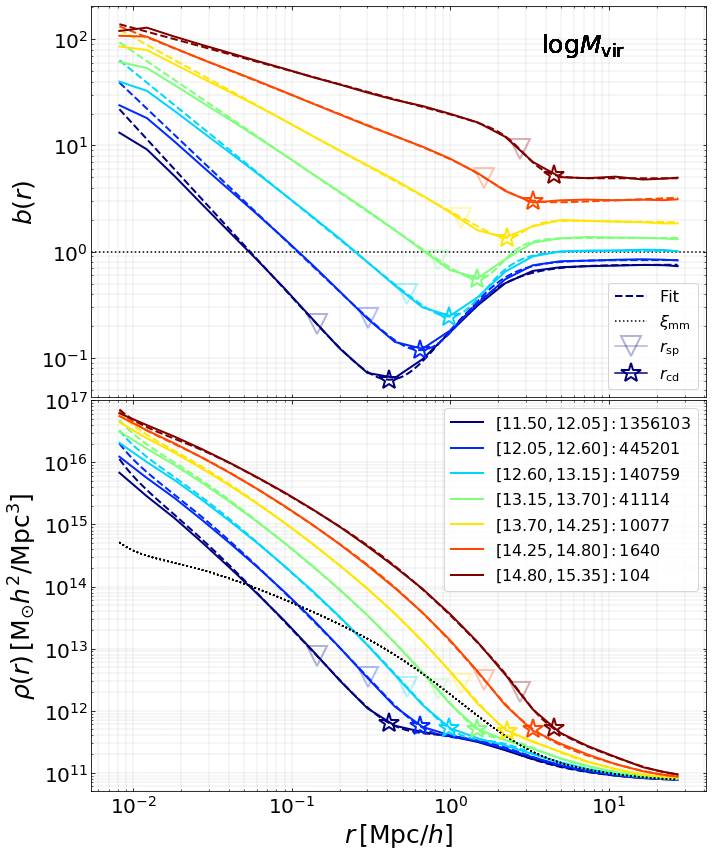}
	\caption{The bias (top panel) and density (bottom panel) binned in separate $\log M_{\rm vir}$ ranges, where each colour corresponds to the mass range and number of haloes in the bin shown in the legend. The solid lines show the mean profiles in each bin. 
	The dashed lines are the best-fits to each profile adopting Equation~\eqref{eq:bFit}. The dotted lines are the matter auto-correlation function expressed as a bias or density in the top or bottom panel, respectively. The stars are the locations of the depletion radii, $\rd$. 
	For reference we plot empty upside down triangles as the splashback radii locations, $r_{\rm sp}$. 
	}
	\label{fig:Mvir}
\end{figure}

With the universal fitting function, we identify the characteristic depletion radius, $\rd$, to be the location of the bias minimum from the fitted bias profile. In Figure~\ref{fig:Mvir} we show the halo bias and density profiles for haloes binned by virial mass, together with the estimated depletion radius. Because our fitting function asymptotes to a single power-law for the one halo bias profile, we limit our fits to start from only the outer part of the one-halo profile at $r \geq 0.06 \, {\rm Mpc}/h$. For the two highest mass bins, as there is not a well defined minimum, we estimate the radius to be where the bias just flattens to the linear bias. For comparison, the splashback radius is also shown for each bin by finding the radius of the steepest logarithmic slope in the fitted density profile, $\rho(r)=\rho_{\rm m} \times (b(r)\xi_{\rm mm} +1)$. \footnote{We also tried identifying the splashback radius by fitting with a Diemer and Kravtsov density profile~\citep{Diemer2014}, and find it does not perform well for all of the forms the density takes on in this work, especially when binning by two halo parameters. Besides, when binning by mass, the density profile fitting function of~\citet{Diemer2018} does not always capture the build up of matter between $\rd$ and the linear bias, but still recovers the mean and median virial masses well.} 

In Figure~\ref{fig:Mvir} the bias is dominated by the 1-halo term for smaller radii, while for larger radii the bias flattens to the expected linear bias. As discussed before, the trough in the intermediate scale can be interpreted as a result of the competing mass accretion between the central halo and neighbouring ones. In the hierarchical structure formation framework, smaller haloes form earlier and thus have well-formed accretion troughs after depleting matter around the halo. On the other hand, the most massive haloes are still actively accreting matter from the neighbourhood with fewer competitors, leading to shallower or even no obvious accretion troughs. 

The transition between $\rd$ and the linear bias for each of the lower mass bins show a positive slope before flattening to the linear bias, which can also be seen as a flat shoulder in the density profiles (Figure~\ref{fig:Mvir}). This shoulder marks a transition scale where the density profile starts to deviate significantly from the inner power-law shape, and again justifies the choice of $\rd$ as a natural choice for halo boundary. The splashback radius, on the other hand, is always located within $\rd$ while the density profile extends naturally beyond $r_{\rm sp}$ out to $\rd$. As we will see later when studying the dynamics, the change in the shape of the density profile reflects the transition from the inner growing halo and the surrounding environment being depleted over time.

\begin{figure} 
	\includegraphics[width=\columnwidth]{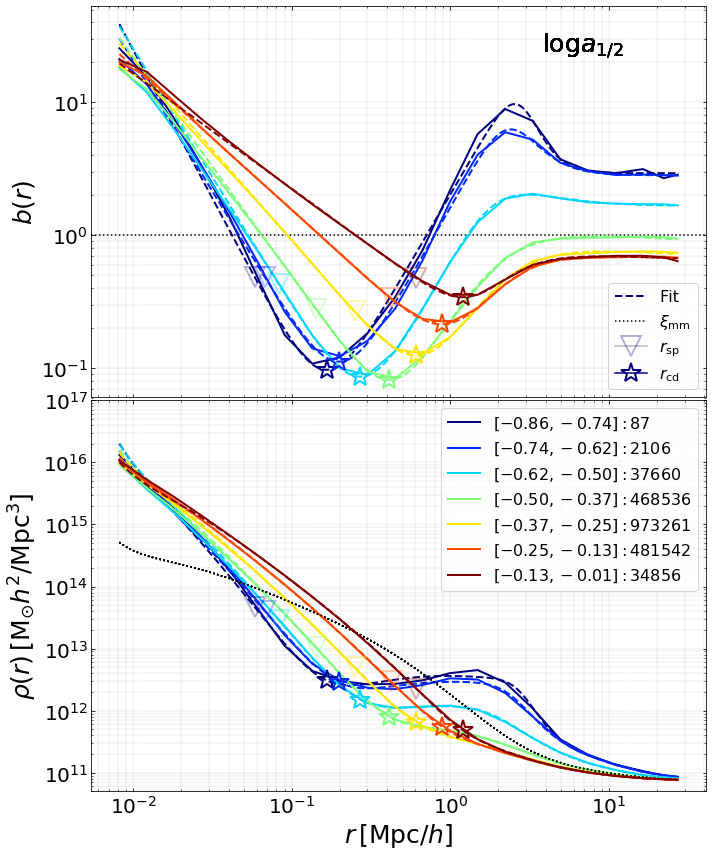}
	\caption{Similar to Figure~\ref{fig:Mvir}, but showing the bias (top panel) and density (bottom panel) binned in separate $\log a_{1/2}$ ranges as labelled. 
	}
	\label{fig:a_half}
\end{figure}

The transition region between the 1 and 2-halo terms can sometimes be more pronounced for the bias when binned by other halo parameters. The bias and density binned by halo formation time can be seen in Figure~\ref{fig:a_half}, where the colours, lines, and symbols are similar to Figure~\ref{fig:Mvir} but in different $\log a_{1/2}$ bins. The locations of $\rd$ are again located at the starting point of the shoulder region in the density profiles. For the earliest forming haloes, the bias profiles also show pronounced peaks outside the trough, reflecting the significance of neighbouring haloes.\footnote{The high biases for the early forming haloes could also be partly contributed by splash-back haloes, i.e., those ejected from bigger haloes and are thus found with massive neigbours. These haloes account for some of the most extreme values of halo properties, such as concentration and spin, and their massive neighbours have significant impacts on the large-scale bias~\citep[e.g.][]{Mansfield2020,Tucci2020}. However, these splash-back haloes are expected to be only a minor part of the early forming population~\citep{Wang2007,Han2019}.} These walls also lead to more extended shoulders in the density profiles. This remains the case for the other parameters listed in Section~\ref{sec:simulation}.%, and the figures can be seen in Appendix~\ref{sec:1DBinning}. 
It is also clear that the trough is well-formed in early forming haloes, while it is shallower in recently formed ones, as the environment are relatively less depleted for the latter.

It is important to note that the motivation of binning the bias profile by a single halo property is to explore its behavior presuming we are ignorant about the primary driver for our new halo boundary definition. However, as we show below, this new boundary of a halo is primarily determined by halo mass among the list of halo properties we investigate. When binning by halo properties besides mass, we are stacking haloes of varying sizes and shapes. 
Furthermore, many of the halo parameters used in this work (Section~\ref{sec:simulation}) are tightly correlated with virial mass. Therefore the observed dependencies of $\rd$ on non-mass parameters could be largely contributed by the correlation between mass and the secondary parameter. However, we include the above figures to help visualize the connection between the bias and density, and show that the characteristic depletion radius represents a meaningful location.
A fairer study of the impact of a secondary parameter would be to bin the bias by both mass and secondary halo property, as detailed below.

\begin{figure*} 
	\includegraphics[width=\textwidth]{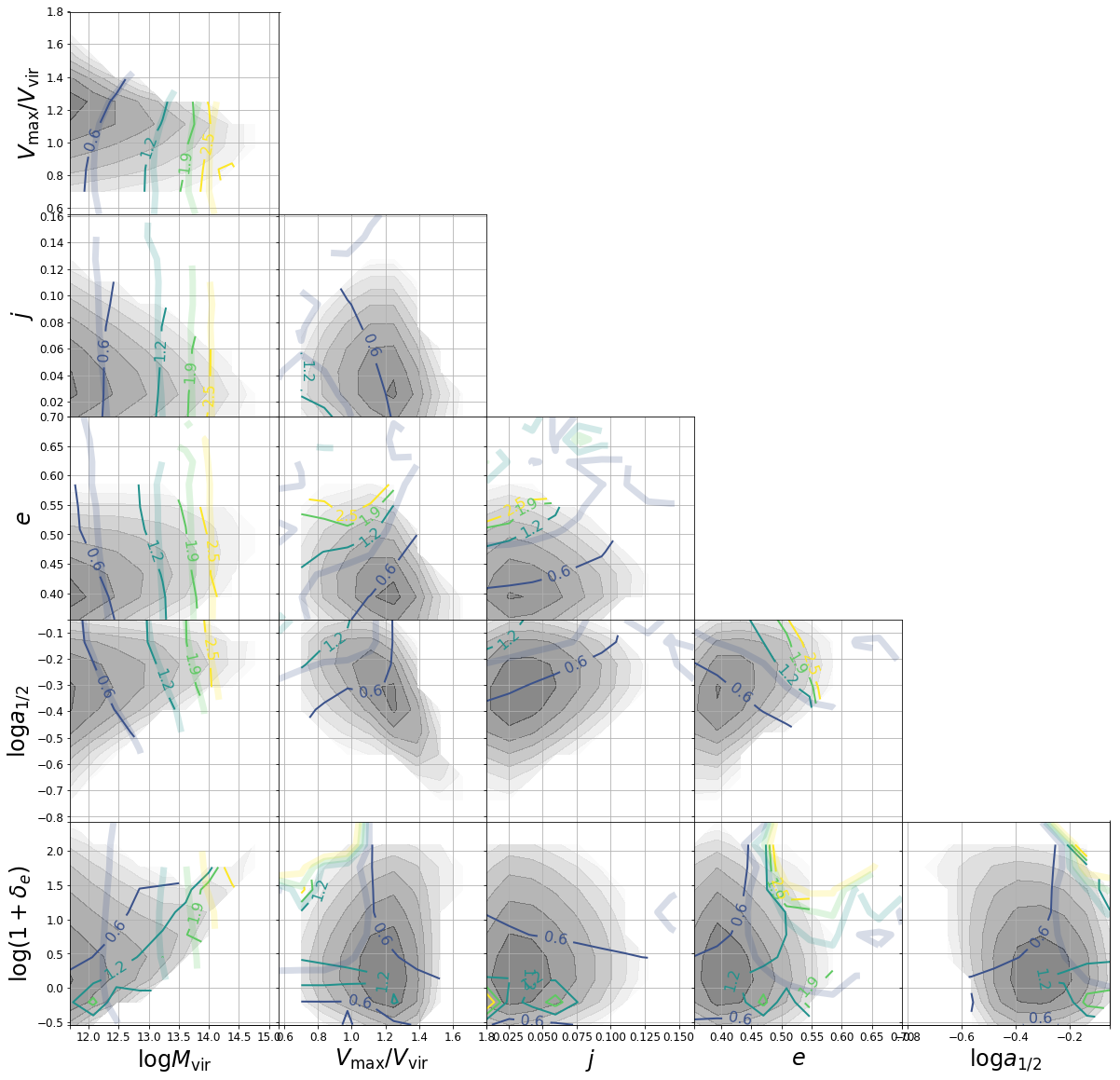}
	\caption{The $\rd$ values for the bias binned by two halo properties, for all combinations in Section~\ref{sec:simulation}. Here we use a total of 10 bins logarithmically spaced for each parameter. The grey filled contours show the number density distribution of haloes for bins containing 100 or more haloes. The thin solid lines are contours of the $\rd$ values, and the thick transparent lines are predictions from a GPR model $\hat{r}_{\rm d}(M_{\rm vir}, a_{1/2})$ trained on the $(\log M_{\rm vir}, \log a_{1/2})$ panel.
	}
	\label{fig:allMaps_Rmin}
\end{figure*}

In Figure~\ref{fig:allMaps_Rmin} we show the $\rd$ values for the bias binned by two halo properties, for all combinations in Section~\ref{sec:simulation}. The grey filled contours are the number of haloes in each bin, which trace the signal-to-noise (S/N) of the measurements. The directions of the gradient of the halo depletion radius tells how sensitive $\rd$ is to the variations of halo properties.
Among all the halo properties except the environment, $\rd$ depends mostly on halo mass. 
This can be seen as the contours mostly changing in the mass direction, when binned by one halo parameter and mass. There are some dependencies on other halo properties besides mass, among which halo spin, $j$, is the least sensitive parameter for $\rd$. For haloes of the same virial mass, early forming ones are smaller in $\rd$, and so are spherical, highly concentrated, and high spin ones. The dependence on the halo environment, $\delta_{\rm e}$, is also significant, with haloes in low density environment being larger, consistent with the picture that the influence range of the halo can be larger in low density environment, while the competition from the environment is stronger when the environment is higher. 

In principle we can further bin the haloes by three or more halo properties to study their joint dependences, but this becomes difficult to visualize once the dimension is above two. Given that the dependence on many halo properties are already weak, in this work we simply test if there are significant residual dependencies on other halo properties once we account for the mass and formation time dependence. This is done by first fitting the $\rd$ dependence on mass and formation time with a flexible function, $\hat{r}_{\rm d}(M_{\rm vir}, a_{1/2})$, and then recast the fitted depletion radius to bins in other halo properties. Following~\citet{Han2019}, the fitting is done through Gaussian Process Regression (GPR), which can be regarded as a flexible non-parametric smooth interpolation of the ${r}_{\rm d}(M_{\rm vir}, a_{1/2})$ map obtained in Figure~\ref{fig:allMaps_Rmin}. 
In our work we use the Gaussian Process Regressor implemented in \textsc{Scikit-Learn}~\citep{Pedregosa2011},\footnote{\url{https://scikit-learn.org/stable/modules/gaussian_process.html}} adopting a Matern kernel with $\nu=0.5$. When fitting the $\rd$ map we bootstrap the sample of haloes in each bin to estimate the noise of $\rd$. The bootstrapped halo depletion radius values have non-normal distributions so we take the noise to be the maximum of the $\pm 34^{\rm th}$ percentile from the median of the distributions as the noise for the GPR fits.

The contours from the GPR fits are overplotted in Figure~\ref{fig:allMaps_Rmin} as thick contour lines. For panels other than the ($M_{\rm vir}, a_{1/2}$), these thick contour lines represents predictions from the GPR fits recasted into the corresponding halo property bins. Unlike the ($M_{\rm vir}, a_{1/2}$) panel, the halo bins in other panels can have wide distributions in mass and formation time. To compute the GPR prediction for these bins, we calculate the mean value of the predicted depletion radius from the GPR, $\langle \hat{r}_{\rm d}( M_{\rm vir},  a_{1/2}) \rangle$, as the recasted depletion radius in each bin, where the average is taken over the distribution of haloes in the bin. Note there is a caveat in this approach as the size for a sample of haloes computed from their stacked profiles can be different from their average size. Ideally the binned radius should be predicted from the stacked theoretical bias profiles, for which a full halo model is needed. Despite this, we expect the average predicted radius to still be able to largely inform us of the variation trend of the model in different bins. As shown in the figure, when binning by 1 or 2 other halo parameters the mean GPR predictions fail to recover the true $\rd$ values, indicating the dependence on the other properties are unlikely to be fully attributable to the mass and formation time dependencies. This is especially true for the dependence on the environmental density.

%%%%% Rd vs Rsp 
\subsection{Relation to the splashback radius}
\label{sec:RbvsRx}

\begin{figure} 
	\includegraphics[width=\columnwidth]{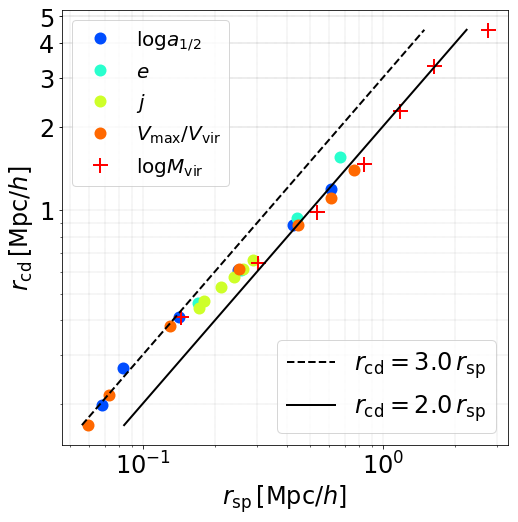}
	\caption{Relation between the characteristic depletion radius, $\rd$, and the splashback radius, $r_{\rm sp}$. These radii are measured from the density and bias profiles in Figure~\ref{fig:ALL_bias}, binned according to various halo properties as shown by the different symbols. The dashed and solid lines are reference linear relations as labelled. }
	\label{fig:RspvsRmin}
\end{figure}
We compare the halo depletion radius, $\rd$, with the splashback radius, $r_{\rm sp}$, in Figure~\ref{fig:RspvsRmin} when the bias is binned by one halo parameter. We use red $+$ symbols to emphasize the mass cases. The dependence of $\rd$ on the environment appears more complicated than the others, which we do not consider here but leave to future investigations, given that the $\delta_{\rm e}$ is not a classical halo property. Besides, we also exclude bins with less than 100 haloes out of S/N consideration, and the shape parameter bin $[0.55, 0.61]$ due to the difficulty in estimating a proper depletion radius for it.

A line $r_{d}/r_{\rm sp} = 2$ provides a very good description of the relation between the two radii, at least for the majority of the halo mass range. The ratio becomes larger at the small radius end, approaching $\rd/r_{\rm sp} \simeq 3$. This can be interpreted as reflecting the higher concentration of these small haloes, which can be further related to the early formation time, analogous to the well-known behaviour of the conventional NFW concentration~\citep[e.g.,][]{Bullock2001,Wechsler2002,Zhao09,Ludlow13} which is also clear from the halo distribution in the $(a_{\rm 1/2}, V_{\rm max}/V_{\rm vir})$ panel in Figure~\ref{fig:allMaps_Rmin}. However, it is interesting to see that this ``outer concentration'' (or puffiness) depends mostly on the halo size and approaches constant values at both the large and small size ends, reflecting the approximately universal shapes of the density profiles near the halo boundary.
As discussed earlier, binning by a single secondary halo parameter can be difficult to interpret due to the tight correlations with halo mass. When binning only by a secondary halo parameter, we are averaging over a wide range of halo sizes. Therefore much of the trends in the depletion and splashback radii above can be driven by their dependence on mass (see Figure~\ref{fig:allMaps_Rmin}). 
However, we include the above example to help guide the reader in visualizing the outer universal shapes in relation to $r_{d}/r_{\rm sp}$, before studying the effects of the secondary parameter along with halo mass.

\begin{figure*}
	\includegraphics[width=\textwidth]{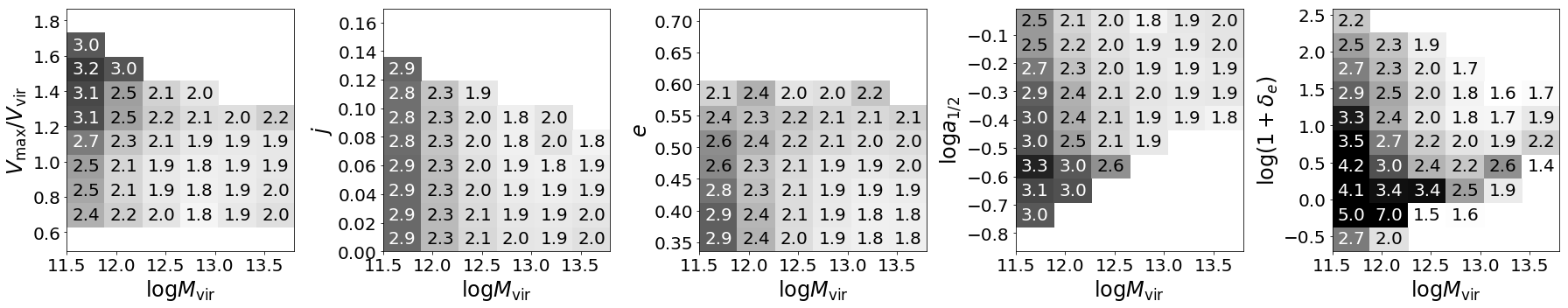}
	\caption{The ratio between the characteristic depletion radius and the splashback radius, $\rd/r_{\rm sp}$, binned by mass and another halo property. Each pixel is colour coded according to the $\rd/r_{\rm sp}$ value in the bin, which is also printed directly.}
	\label{fig:allMaps_Rsp_Rmin}
\end{figure*}

We explore the outer concentration further in Figure~\ref{fig:allMaps_Rsp_Rmin} to study its joint dependence on mass and another halo property. We focus on combinations including mass, as mass is shown to be the most sensitive internal halo variable (i.e., excluding $\delta_{\rm e}$) for the depletion size according to Figure~\ref{fig:allMaps_Rmin}. For combinations of internal halo properties the ratios are mostly in the same range of $\rd/r_{\rm sp}\sim 2-3$ as found before. Note that we do not include the highest mass bins due to the difficulty in confidently estimating the halo depletion radius where there is no trough, similarly to the highest mass bin in Figure~\ref{fig:Mvir}.
This figure shows that for most haloes the splashback radius has a similar trend to the halo depletion radius. Though the differences here are small, it is good to see that the ratio is not constant in different parts of the parameter space, which means $\rd$ carries extra information about the halo properties besides those already contained in $r_{\rm sp}$. 

For the joint dependence on mass and other internal halo properties, the global dependence on mass is still the most significant at the low mass end. The dependence on other halo properties are also significant except for the little dependence on halo spin. The earliest forming, least massive, spherical, and most concentrated haloes also have high outer concentrations, qualitatively tracing the behaviour of the inner NFW concentration.

When the environment is involved, haloes with a small $\delta_{\rm e}$ tend to have a higher outer concentration. This can be understood as the halo boundary becomes more extended in a low density environment as seen in Figure~\ref{fig:allMaps_Rmin}. However, we warn that $\rd$ may be unreliably estimated for the very lowest environmental overdensity bins with $\log (1+\delta_{\rm e}) \lesssim 0.24$.

This type of analysis could be an interesting starting point for a deeper study into the shapes of halo profiles using different boundary definitions, which we leave to future work.

\section{Dynamical interpretation of the depletion trough}
\label{sec:Rd_velocityComparison}
\begin{figure} 
	\includegraphics[width=\columnwidth]{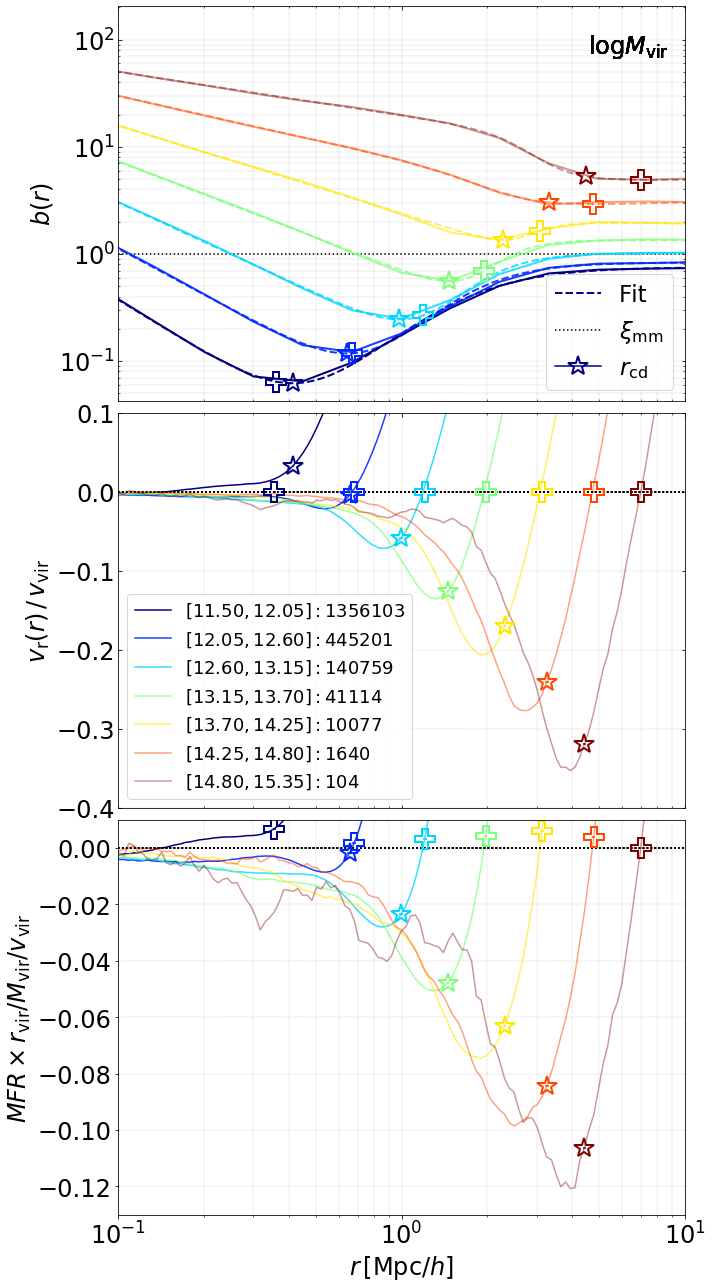}
	\caption{The stacked radial velocity and mass flow rate profiles binned by virial mass. The colours in the plot represent the mass bins in the legend. The middle and bottom panels show the total radial velocities and mass flow rates, respectively. For comparison the top panel reproduces the bias profiles from Figure~\ref{fig:Mvir}. The star and cross symbols mark the characteristic depletion radius and the turnaround radius respectively.
	}
	\label{fig:Mvir_velocity}
\end{figure}

To get more physical insights into the meaning of the depletion radius we study whether $\rd$ corresponds to any features in the dynamics of matter near the halo boundary. As the halo mass is the primary driver (except $\delta_{\rm e}$) for the characteristic depletion radius, we will focus on binning by halo mass in this section.

\subsection{The inner depletion radius revealed in mass flow rate}
In Figure~\ref{fig:Mvir_velocity} we plot the stacked bias, radial velocity, and mass flow rate (MFR) profiles, for haloes binned by mass. The velocity profiles are the mean of the average velocity profiles of the haloes in each bin. The total radial velocity can be decomposed as $v_{\rm r} = v_{\rm p} + v_{\rm H}$, where $v_{\rm p}$ and $v_{\rm H}$ are the peculiar and Hubble velocities, and a negative velocity means infalling motion towards the halo centre. 
The mass flow rate is the amount of mass flowing through a cross-sectional area per unit time, ${\rm MFR} = \rho(r) \, v_{\rm r}(r) \, 4 \pi r^{2}$, and is scaled by $M_{\rm vir} \, v_{\rm vir}/r_{\rm vir}$. The virial velocity is $v_{\rm vir}=\sqrt{GM_{\rm vir}/r_{\rm vir}}$, computed using the median virial mass and the corresponding radius in each bin. According to the continuity equation, 
\begin{equation}
4\pi r^2 \dot{\rho}\ud r=-\ud {\rm MFR},
\end{equation}
the MFR profile determines the evolution of the density profile.

As shown in Figure~\ref{fig:Mvir_velocity}, the MFR and the velocity profiles have very similar shapes. For massive haloes the surrounding matter is being actively accreted onto the haloes, as shown by the prominent infall velocity and mass flow troughs. The lack of an infall region for the lowest mass bin can be interpreted as low mass haloes having completed their accretion phase and that the combination of tidal forces from the neighbouring haloes and the Hubble flow becomes relatively more significant and overcomes the gravity of the main haloes. Separating these low mass haloes according to their environments could give us a better insight into this feature. However, this is not crucial for this current paper and we leave it to a future work.

For the mass bins with velocity troughs, the locations of maximum infall are nearly identical with those of the maximum mass inflow rates. This is because the density profiles are close to $\rho(r)\propto r^{-2}$ around these scales, so that ${\rm MFR}(r)\propto v_{\rm r}(r)$. Because the MFR is the intrinsic quantity determining the evolution of the density profile, we will call this location the maximum inflow radius to emphasize its connection to the density profile evolution.

\begin{figure} 
	\includegraphics[width=\columnwidth]{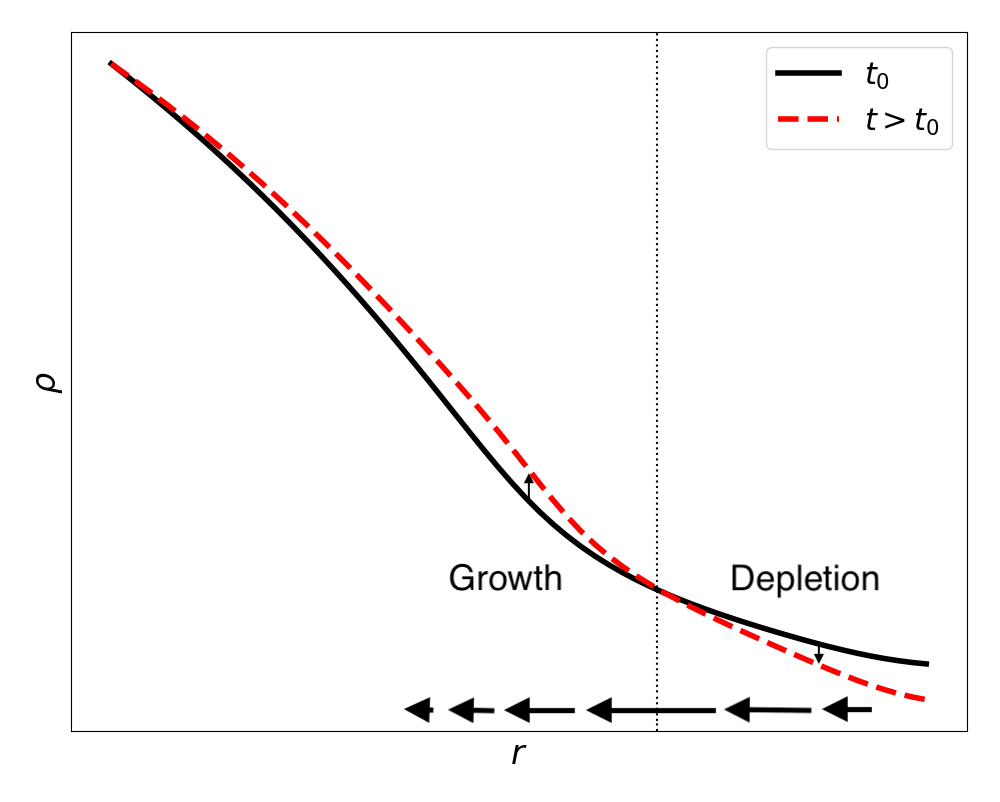}
	\caption{Illustration of halo accretion around the maximum inflow location. The maximum inflow location or inner depletion edge is represented by the vertical dotted line, while the horizontal arrows represent the mass flow magnitudes around the maximum inflow location (see Figure~\ref{fig:Mvir_velocity}). The evolution of the halo is represented by the solid black to the dashed red line. The density profile around the halo grows within the maximum inflow radius, but decreases outside it.
	}
	\label{fig:haloAccretionIllustration}
\end{figure}

Within the maximum inflow radius, matter is being dumped onto the halo as the MFR slows down towards the inner halo. Outside this radius, however, matter is being pumped into the halo and gradually depleted over time due to the increasing infall rate with decreasing halo-centric distance. We provide an intuitive illustration of this process in Figure~\ref{fig:haloAccretionIllustration}. This maximum inflow location thus marks the transition between the halo being built up and matter being depleted by halo accretion in an expanding Universe, leading to the formation of the trough in the bias profile and the flattened shoulder in the density profile. In a follow-up paper, we will show this more explicitly by studying the evolution of the bias and density profiles. Note the halo boundary is expected to grow as the halo grows.

According to this picture, the maximum inflow radius marks \emph{the inner edge} of the active depletion region around the halo. We thus expect the characteristic depletion radius, $\rd$, defined at the minimum bias location to be outside this maximum inflow radius, as $\rd$ is the location where the bias profile is depleted the most. This is indeed the case in Figure~\ref{fig:Mvir_velocity}, where $\rd$ is larger than the maximum inflow radius by $\sim 10-20\%$. With this dynamical interpretation, we can collectively call the maximum inflow radius as an ``inner depletion radius", $\rid$, that marks the starting point of ongoing depletion, while the minimum bias radius as a characteristic or ``deepest depletion radius", both of which characterize the scale of the depletion trough. This bias trough is well-formed around low mass haloes that have completed their mass accretion, but weak around massive haloes that are still in the early stage of their mass accretion.

\subsection{Comparison with the turnaround radius}\label{sec:turnaround}
The turnaround radius can be found where the radial infall of the particles around a halo is just overcome by the Hubble flow. The dynamical interpretation of the depletion radius explained above generally requires $\rd$ to be located within the infall region that is bounded by the turnaround radius in order for the depletion to be caused by accretion. This upper bound also means depletion due to halo accretion only happens in a finite radial range, such that the large scale behavior is not significantly affected by the halo accretion and approaches the average large-scale growth rate in the Universe, resulting in a flat large-scale bias profile and a depletion trough around the boundary scale. Thus the turnaround radius can be interpreted as the ``outer edge" of the active depletion region.

Indeed $\rd$ is always located within the turnaround as shown in Figure~\ref{fig:Mvir_velocity} except for the lowest mass bin.
Note that the total radial velocity remains positive on small scales for the lowest mass bin, likely due to tidal stripping from neighbouring haloes. In this case the turnaround radius is ill-defined, and we estimate it to be the upturning point from the relatively flat inner velocity profile. This ill-behavior could be potentially removed adopting our new halo size definitions in halo finding, so that some of the low mass haloes close to massive neighbours would not be identified as distinct haloes any more. We defer such a self-consistent study to future works.

For high mass haloes, it is expected that the turnaround radius can further grow as the cluster grows more massive. $\rd$ is also well within the turnaround radius, leaving enough space for the halo to accrete from. For low mass haloes, however, the turnaround radius approaches $\rd$. Consequently, the active depletion trough also narrows towards the low mass end, leaving little space for the haloes to accrete from, so that they are barely growing on turnaround scales~\citep[see also][]{Prada2006}. 
This is consistent with the finding of~\cite{Tanoglidis2015}, that haloes below the transitional mass scale of $\approx 10^{13} \, {\rm M_{\odot}}$ have reached their maximum turnaround radius. 

We do not include the 2D binning analysis to compare the depletion and turnaround scales due to significant noise in the velocity profiles. However, we wish to simply show the potentiality of using these two scales in concert with each other to gain insight on the accretion phases of haloes. We leave a more detailed study to a future work.

\begin{figure*} 
	\includegraphics[width=\textwidth]{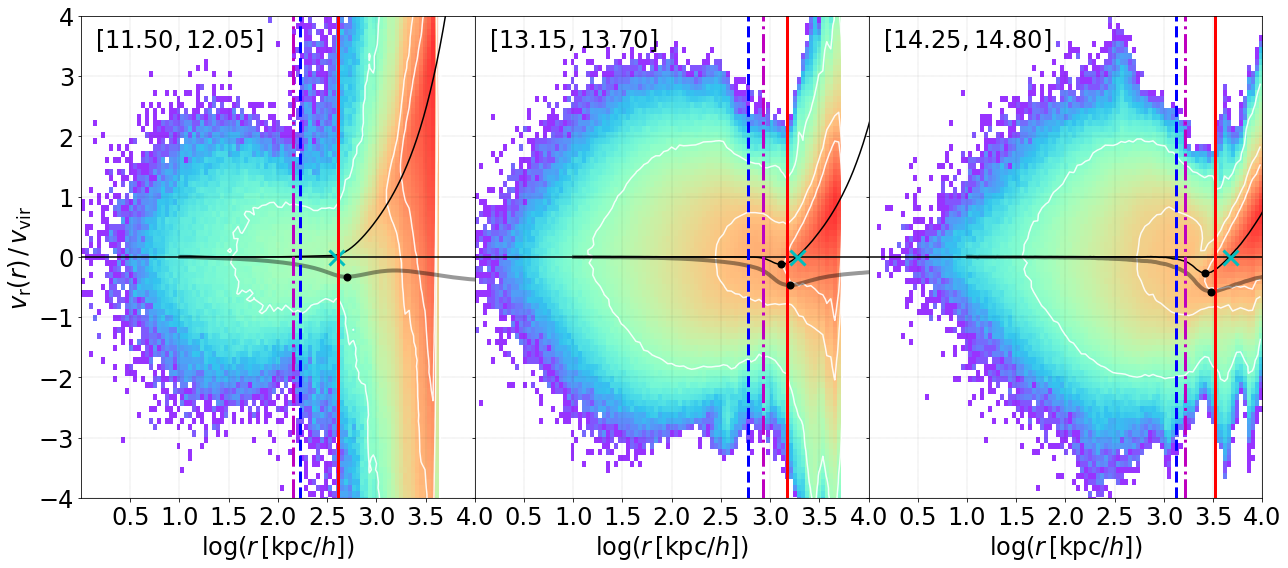}
	\caption{The scaled radial velocity distribution for three $\log M_{\rm vir}$ bins as labeled on the top left of each panel, for 100 randomly selected haloes in each bin. The colour map shows the distribution particles, with white curves marking isodensity contours enclosing 99, 80 and 60 percent of particles from outside to inside. The black and grey curves are the radial and peculiar velocities, respectively. The thick vertical solid red, dash-dotted magenta, and dashed blue lines are the locations of the depletion, splashback, and virial radii for the original full sample in each bin, respectively. 
	The cyan cross symbols are the locations of the turnaround radii, and the points mark the maximum infall locations.
	}
	\label{fig:Mvir_vr_distribution}
\end{figure*}

\begin{figure} 
	\includegraphics[width=0.5\textwidth]{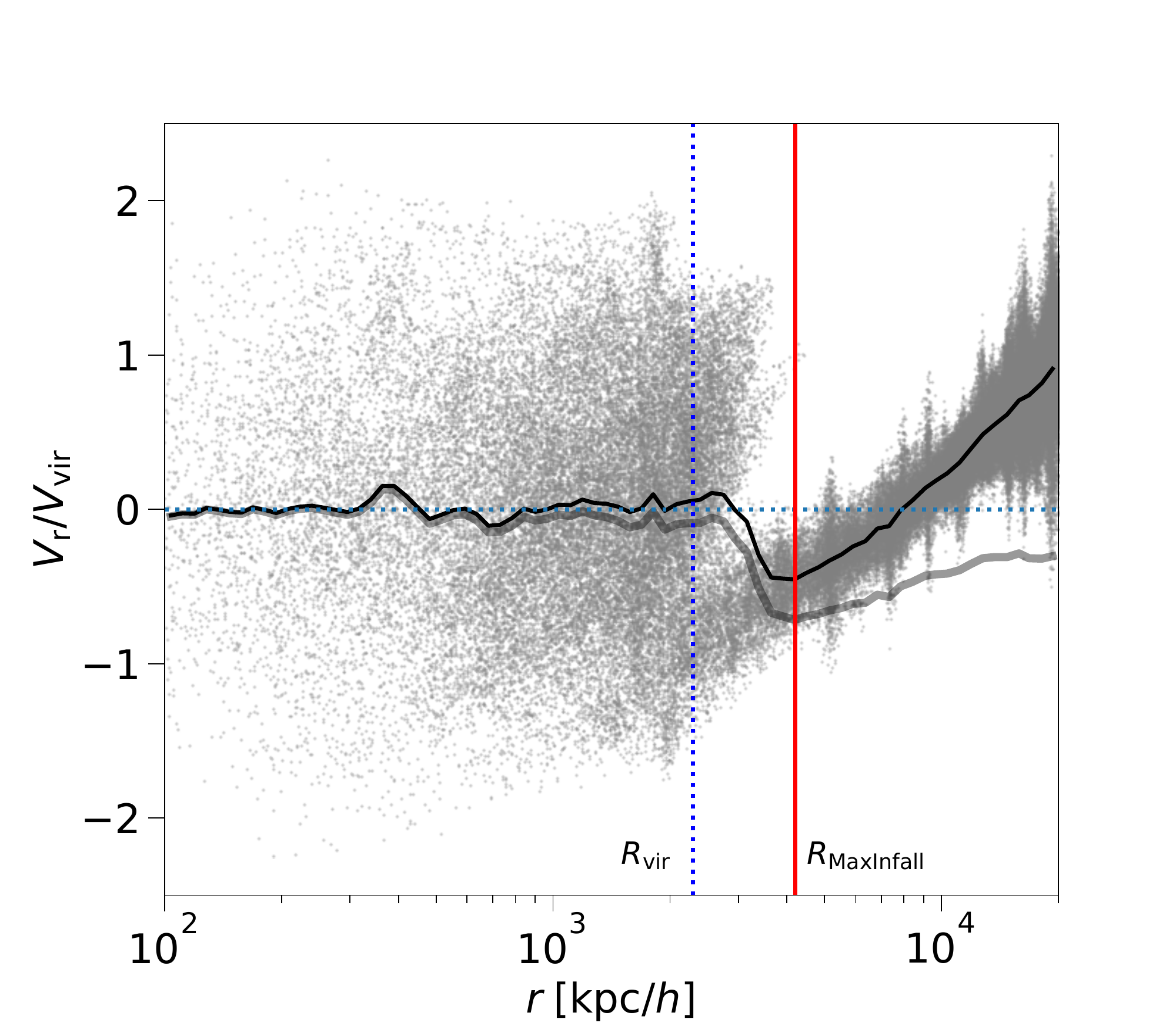}
	\caption{The radial velocity distribution in a single cluster halo ($M_{\rm vir}=1.35\times 10^{15}{\rm M}_\odot/h$). The grey dots show the distribution of halo particles for which only a random 1 percent of all the particles are shown. The black and grey solid curves show the average total and peculiar radial velocities, respectively. The two vertical lines mark the location of the virial and the maximum infall velocities as labelled. The maximum infall location clearly marks the outermost splashback boundary.}
	\label{fig:vr_distribution_single}
\end{figure}

\subsection{Interpretation from particle orbits}

To further demonstrate the significance of the depletion trough in the dynamics around haloes, we show the phase space distribution of particles in haloes of three example mass bins in Figure~\ref{fig:Mvir_vr_distribution}. Note for the distributions we have only stacked a random subset of 100 haloes in each mass bin. The velocities are scaled by the virial velocity computed using the median virial mass and the corresponding radius in each bin. The Hubble flow has been included in the velocity. 

There are obviously two distinct components in the phase space diagram, belonging to the halo and the surrounding environment respectively. Overall the splashback radius, the depletion radii and turnaround radius are all located around the boundary separating the two components. The depletion radii can be found as marking the place where the phase space distribution is the narrowest according to the density contours. 

The median peculiar radial velocity profile is also shown in each panel. Intriguingly, $\rd$ appears to be closer to the location of the maximum peculiar infall velocity, both of which are slightly outside the radius of the maximum total infall velocity. The peculiar velocity profile determines the evolution of the density profile in comoving coordinates through the comoving mass flow rate. However, this maximum peculiar infall location does not correspond to a peak in the comoving mass flow rate, preventing us from making a physical connection between the density profile evolution and the maximum peculiar velocity location.

In terms of the splashback radius, the inner depletion radius, $\rid$, can be roughly interpreted as the outermost splashback radius enclosing the complete population of splashback orbits, or at least enclosing a much higher fraction than the steepest slope radius which is found to be enclosing approximating 75\% of the splashback orbits~\citep{Diemer2017B}. As shown in Figure~\ref{fig:Mvir_vr_distribution}, the steepest slope splashback radius is also not far from the boundary of the narrowest velocity distribution, but further inside. To show the splashback interpretation of the depletion radius more clearly, in Figure~\ref{fig:vr_distribution_single} we show the phase space distribution of a single cluster halo. Outside the halo boundary, the velocity distribution is dominated by the infalling component, with an increasing infall velocity as material falls closer to the halo. Once the accretion stream enters the halo boundary, the phase space will also be filled by splashback orbits or ejected particles that contribute to outflowing velocities, leading to an decrease in the net infall velocity. As a result, the location of the maximum infall marks the apocenter of the outermost splashback orbits. This is not as clear in the stacked phase space diagrams in Figure~\ref{fig:Mvir_vr_distribution}, potentially due to the mixing of haloes of different size and the accumulated blurring from fluctuations around the boundary when many haloes are stacked.

The difference between the splashback radius defined at the steepest slope and that defined through the infall velocity profile may also be understood considering the asphericities of haloes. Because haloes are typically aspherical and because there will be halo-to-halo scatter~\citep[e.g.][]{Mansfield2017,Diemer2017A}, the steepest slope radius can be interpreted as a spherical average of the typical splashback radii in the given halo sample. This average radius is expected to be smaller than the radius encompassing the complete population of splashback orbits.

A detailed statistical comparison of the depletion radii with the high percentile splashback radius would be interesting but not so straightforward due to the difficulty in robustly estimating the splashback radius at the tail of the splashback distributions, and we leave such investigations to future works. The current picture suggests the maximum inflow radius, or inner depletion radius, could be used as an alternative measure of the complete splashback radius.

\begin{figure} 
	\includegraphics[width=\columnwidth]{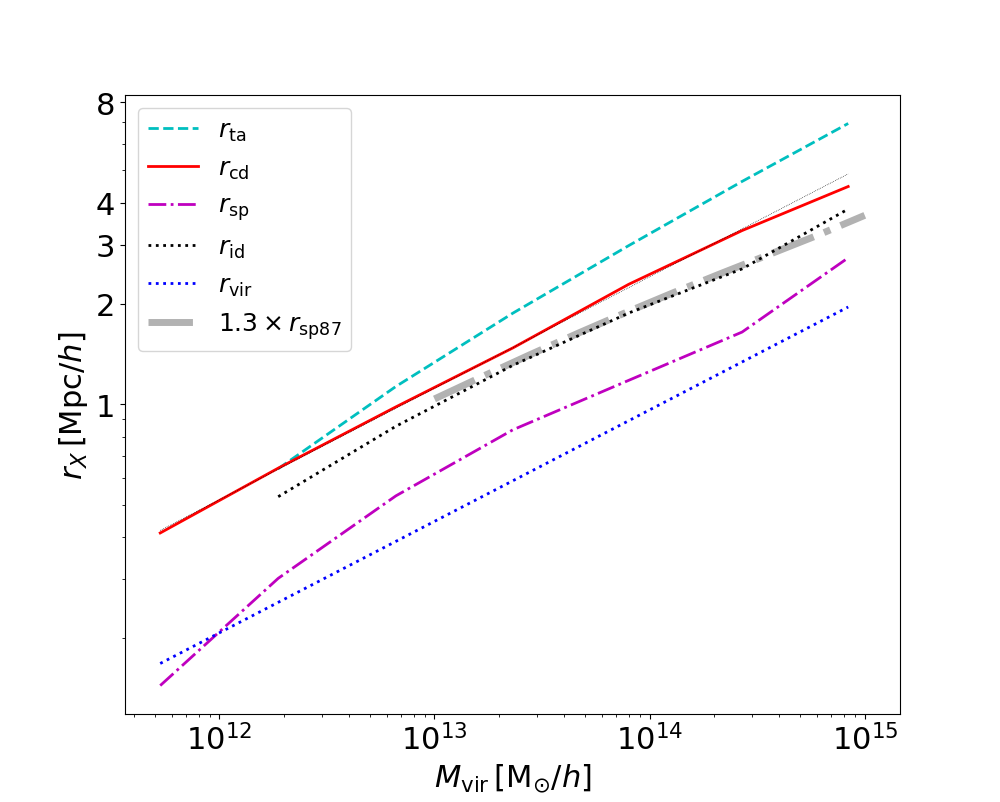}\\
	\includegraphics[width=\columnwidth]{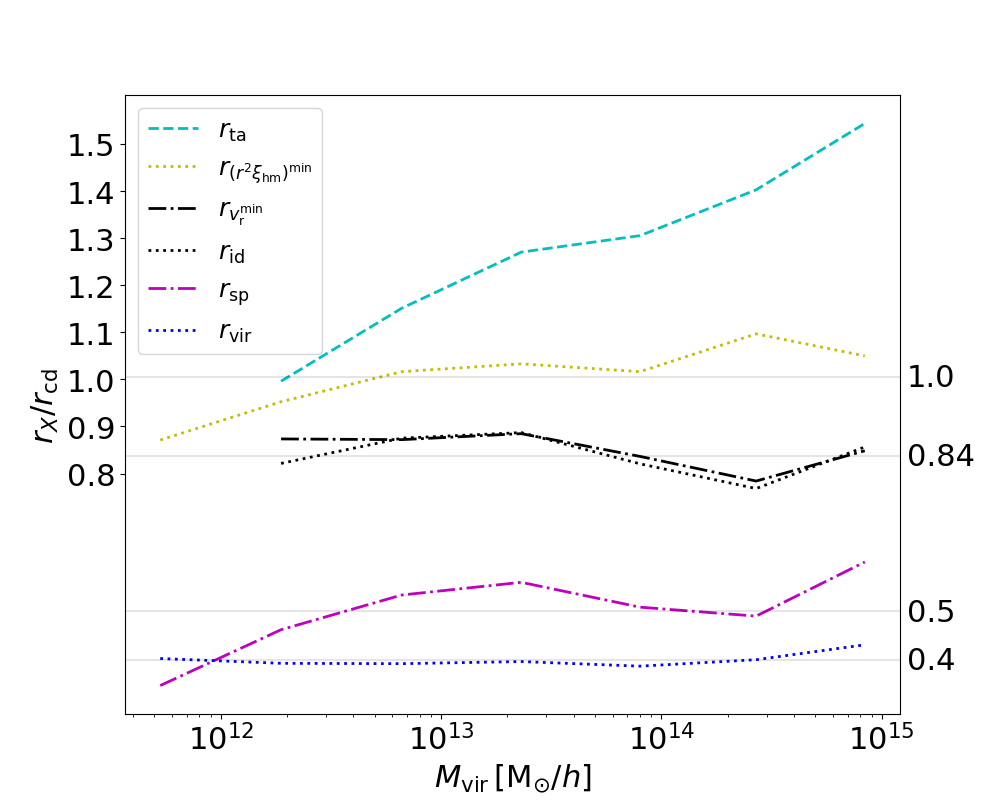}
	\caption{\textit{Top}: The characteristic radii as functions of virial mass. The coloured lines correspond to the radii in the legend. The grey dotted line shows $2.5 r_{\rm vir}$ to compare to $\rd$. The $1.3 r_{\rm sp87}$ radius is a proxy of the optimal exclusion radius found by~\citet{Garcia2020} discussed in section~\ref{sec:haloModel}. \textit{Bottom}: The ratio of the various radii to the characteristic depletion radius $\rd$. In addition to those listed in the top panel, we also include the radius of the maximum in the total infall velocity, $r_{v_{\rm r}^{\rm min}}$, and the radius of the minimum in $r^{2}\xi_{\rm hm}$ (see section~\ref{sec:haloModel}). The horizontal lines are the constant best-fit to the ratios, with their values shown on the right side. We do not include a best-fit to the turnaround radius due to it's strong mass dependence.}
	\label{fig:Mvir_RxVsMvir}
\end{figure}

\begin{figure*} 
	\includegraphics[width=\textwidth]{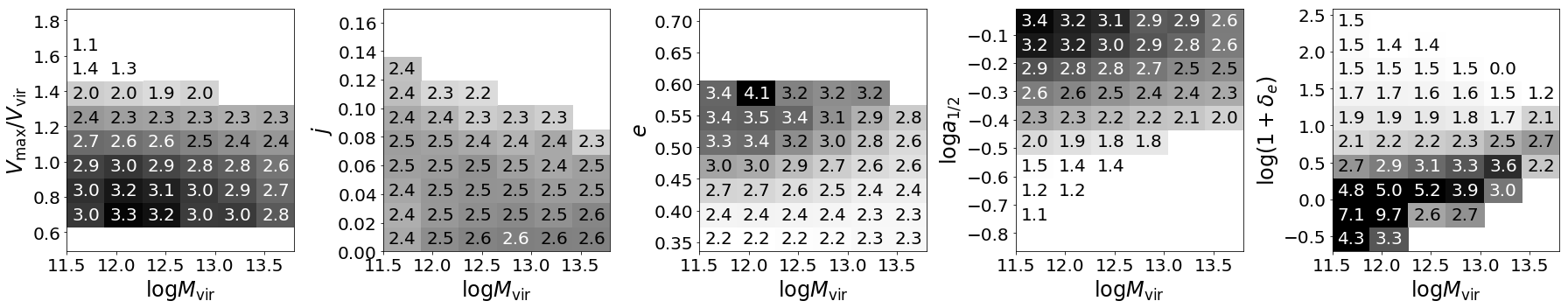}
	\caption{The dependence of the characteristic depletion radius to virial radius ratio, $\rd/r_{\rm vir}$, on combinations of different halo properties. The radius ratio is shown by the pixel value and a corresponding colour. 
	}
	\label{fig:allMaps_Rvir_Rmin}
\end{figure*}

\section{Mass-radius relations and the enclosed density}\label{sec:M-r}

In Figure~\ref{fig:Mvir_RxVsMvir} we compare the various characteristic radii as functions of halo virial mass. As discussed above, the inner depletion radius is close to the characteristic depletion radius, and both are close to the turnaround for haloes below $M_{\rm vir}\sim10^{\rm 13}{\rm M}_\odot/h$. 
The splashback radius, on the other hand, is typically close to but larger than the virial radius. 

We do not include the lowest mass bin for the maximum inflow and turnaround locations here, due to the lack of an infall region in the corresponding $v_{\rm r}$ profile (see the discussion in Figure~\ref{fig:Mvir_velocity}). We also emphasize that $\rd$ for the high mass bins may not be reliably estimated due to their weak or absent trough feature in the bias profile, and some more careful studies are needed to accurately measure the behavior in these mass ranges. We show the ratio of many measured radii to the characteristic depletion radius in the bottom panel of Figure~\ref{fig:Mvir_RxVsMvir}. As we discussed before, the maximum mass inflow rate location is almost identical with the maximum infall velocity location, both of which are lower than $\rd$ by $10-20\%$ in the mass range studied.

The characteristic depletion radius has a constant ratio to the virial radius when binned by halo mass, with the fitted line of $\rd = 2.5 r_{\rm vir}$ in Figure~\ref{fig:Mvir_RxVsMvir} (or $\rd \approx 2.0 r_{\rm 200m}$, for reference). However, the relation becomes more interesting when further binning by other halo parameters. This can be seen in Figure~\ref{fig:allMaps_Rvir_Rmin}, where the ratios $\rd/r_{\rm vir}$ vary with halo properties other than mass when binned by combinations of internal halo properties. An outlier is again the halo spin, which barely affect the ratio. The trends in $\rd/r_{\rm vir}$ are significantly different from those in Figure~\ref{fig:allMaps_Rsp_Rmin}, showing that the various halo radii can probe different properties of haloes. The dependence when $\delta_{\rm e}$ is involved becomes much different from other panels, but roughly consistent with the behavior in Figure~\ref{fig:allMaps_Rsp_Rmin}, reflecting that the $\delta_{\rm e}$ mostly influences $\rd$ rather than $r_{\rm sp}$ or $r_{\rm vir}$. 

Comparing this figure to Figure~\ref{fig:allMaps_Rsp_Rmin}, we find that low $\rd/r_{\rm vir}$ ratios for highly concentrated or early forming haloes correspond to high ratios in $\rd/r_{\rm sp}$, or high ratios in $r_{\rm vir}/r_{\rm sp}$. These correspond to the older, low-mass, highly concentrated, low spin, and spherical haloes (see Figure~\ref{fig:allMaps_Rmin}). As discussed previously, the low-mass haloes are more likely to have completed their accretion phases, and part of the low-mass population can be made up of distinct post-pericentre splashback haloes, or ``ejected subhaloes''. The ejected subhaloes are thought to be one of the causes of assembly bias for low-mass haloes~\citep{Dalal2008,Sunayama2016,Mansfield2020}, where the tidal gravitational forces from their massive neighbours can have significant impacts on their internal properties~\citep{Mansfield2020,Tucci2020}. Seeing how ejected subhaloes can have such incredible impacts on halo properties and bias, it would be interesting to trace the evolution of the sample of haloes in the context of Figure~\ref{fig:allMaps_Rmin}, which we leave to a future work.

\begin{figure} 
	\includegraphics[width=\columnwidth]{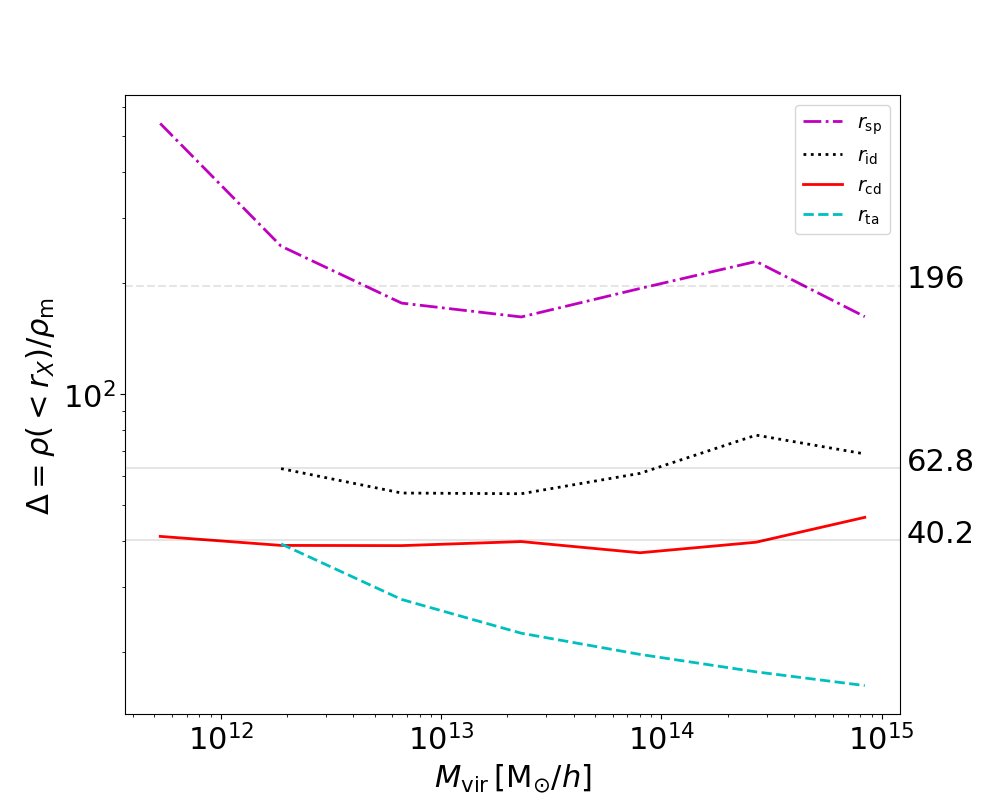}
	\caption{The average density enclosed in different characteristic radii ($r_{X}$ as labelled in the legend) around haloes of different virial masses, normalized by the mean density of the Universe.
	}
	\label{fig:Mvir_DeltaVsMvir}
\end{figure}

In Figure~\ref{fig:Mvir_DeltaVsMvir} we plot the average density enclosed within each characteristic radii, for haloes of different virial masses. The average density within $\rd$ is consistent with a constant of $\rho(<\rd) \approx 40.2\rho_{\rm m}$, independent of halo mass. In other words, this depletion radius can be found as a simple spherical overdensity radius containing 40.2 times the background density (or equivalently $\sim 10.8$ times the critical density) of the Universe in our simulation, when only the mass dependence is considered. Similarly, $\rid$ can also be found enclosing an almost constant density of $\rho(<\rid) \approx 62.8\rho_{\rm m}$ over the mass range probed. By contrast, the splashback radius corresponds to varying overdensities and can not be simply defined through an characteristic overdensity criterion, as also found previously~\citep{Diemer2017A}. This is mostly due to the sharp increase in density at the low mass end. Excluding the lowest mass bin leads to an average enclosed density of $\rho(<r_{\rm sp})\approx 196\rho_{\rm m}$. For completeness, the average density enclosed in the turnaround radius is also shown. It increases with decreasing mass and becomes close to the depletion density at the low mass end. The higher enclosed density at the low mass end can be understood as their turnaround radii froze at higher redshifts where the density of the Universe was higher.

\begin{figure*} 
	\includegraphics[width=\textwidth]{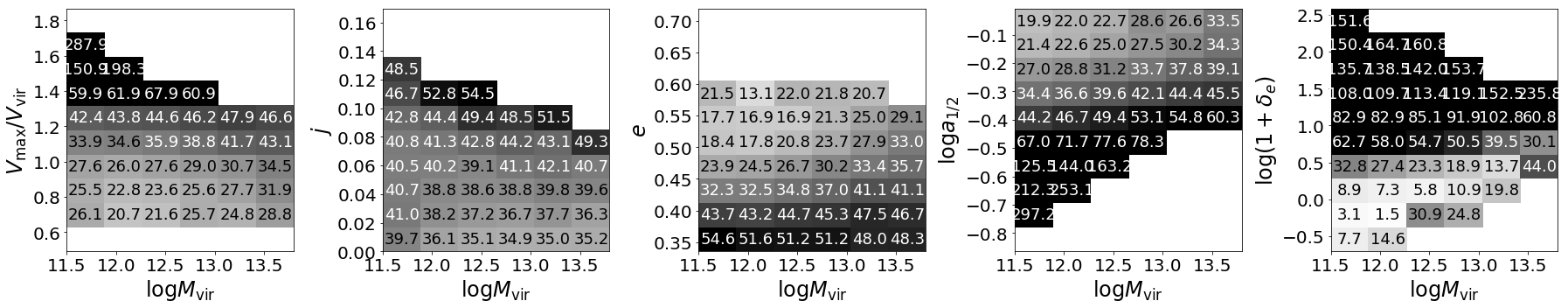}
	\caption{The density contrast, $\rho(<\rd)/\rho_{\rm m}$, when binned by two halo parameters. It is approximately independent on mass but depends on other halo properties. 
	}
	\label{fig:allMaps_Delta}
\end{figure*}

In Figure~\ref{fig:allMaps_Delta} we plot the spherical density, $\Delta=\rho(<\rd)/\rho_{\rm m}$, for the halo depletion radius binned by two halo parameters, one being mass. 
Similar to the behavior of $r_{\rm sp}/r_{\rm vir}$, when we bin haloes by two halo parameters, we find that $\Delta$ is not universal but depends on halo properties other than mass. At a fixed mass, the density variation is largely consistent with the $\rd$ variation in the parameter space, where a smaller $\rd$ corresponds to a higher density contrast. The oldest, spherical, and the most concentrated haloes typically have the highest enclosed density within $\rd$.

Note that combining the radius ratio in Figure~\ref{fig:allMaps_Rvir_Rmin} and the density contrast in Figure~\ref{fig:allMaps_Delta}, one can immediately estimate the ratio between the mass enclosed within $\rd$ and the virial mass. For example, taking the typical $\rd = 2.5 r_{\rm vir}$ and $\rho(<\rd)=40.2\rho_{\rm m}$, the enclosed mass can be found to be $M(<\rd) \approx 1.7 M_{\rm vir}$.

\section{Implication for Halo Model}
\label{sec:haloModel}

We have discussed that the depletion trough can be understood as a manifestation of halo exclusion, such that the radii characterising the trough are expected to find direct applications in the halo modelling of the large-scale structure. A better definition and characterisation of the halo is expected to improve both in the simplicity and the accuracy of the resulting halo model. During the preparation of this work, a very recent outstanding work by~\citep[][hereafter G20]{Garcia2020} has just addressed this problem from an inverse approach compared to ours, by solving for an optimal halo radius definition while optimizing the halo model fitting to the halo matter correlation function. We thus take the leisure to just compare our results to their inferred optimal halo exclusion radius, to demonstrate the significance of our new radius for halo modelling.

In G20, the halo density profile and the optimal halo radius are parametrized through scaling relations with the optimal halo mass. The optimal mass and radius are expected to be different from the conventionally defined ones, so the parameters determining the scaling relations are allowed to vary freely and are solved for by matching the predicted halo-matter correlation to the measurement from simulations. The resulting best-fit scaling relation then provides the basis for defining the optimal mass and radius, by self-consistently interpreting the halo mass as the mass enclosed within the optimal halo radius.

G20 found that their optimal halo radius has a roughly constant ratio to the splashback radius, $r_{\rm G20}\approx 1.3 r_{\rm sp87}$, where $r_{\rm sp87}$ is the radius containing 87\% of the splashback apogees of all the halo particles according to~\citet{Diemer2017B}. To compare against the G20 result, we will use $1.3 r_{\rm sp87}$ as a proxy of $r_{\rm G20}$, where $r_{\rm sp87}$ can be computed from scaling relation with the virial mass. As shown in Figure~\ref{fig:Mvir_RxVsMvir}, for the mass range covered by the G20 model, the G20 radius well matches the inner depletion radius, $\rid$. Note that we have not adopted any tuning in defining our inner depletion radius. Rather our definition comes from complete physical intuition. The excellent agreement between our ad-hoc characterisation of the halo radius and the optimized exclusion radius required by the halo model is thus truly encouraging.

This agreement is also consistent with the phase space interpretation that $\rid$ marks the outermost radius of the splashback particles, or the outer edge of particles orbiting in the halo. The good match indicates this high completeness splashback radius is approximately proportional to $r_{\rm sp87}$. In fact, an apparently better proportionality between $\rid$ and the steepest slope radius, $r_{\rm sp}$ can be found as $\rid\simeq 1.6 r_{\rm sp}$ as seen in Figure~\ref{fig:Mvir_RxVsMvir}.

It was shown in G20 that $r_{\rm G20}$ is located at the minima of $r^{2}\xi_{\rm hm}$ for haloes around $10^{13}\mathrm{M}_{\odot}/h$, which is the "by-eye" boundary of the halo. However, their $r_{\rm G20}$ starts to deviate from the "by-eye" boundary in higher mass haloes. As shown in the bottom panel of Figure~\ref{fig:Mvir_RxVsMvir}, our $\rd$ defined at the minimum bias location is also located very near the minima of $r^{2}\xi_{\rm hm}$ for all mass bins. This is because the matter auto-correlation, $\xi_{\rm mm}$, is found to have a logarithmic slope close to $-2$ around these scales. As a result, the minimum bias location is also expected where $\xi_{\rm hm}$ has the same slope of $-2$, or where $r^2\xi_{\rm hm}$ reaches an extreme. As $\rd$ tends to be further away from $\rid$ in high mass haloes, this explains the poorer agreement between the optimal exclusion radius and the "by-eye" boundary in more massive haloes. However, instead of using the $r^{2}\xi_{\rm hm}$ to define the ``by-eye" halo boundary, we believe that our definition of $\rd$ using the bias profile is more physically meaningful as discussed in section~\ref{sec:binnedBias}.

\section{Summary and Conclusions}
\label{sec:conclusions}

In this paper we have proposed two new characterisations of the halo boundary using a large sample of haloes from a cosmological $N$-body simulation, using the spatial and velocity distribution of matter around haloes. 
In the spatial distribution, the ``characteristic depletion radius", $\rd$, is most evident from the existence of a ubiquitous trough in the bias profile around this halo boundary, where the clustering around the halo is the weakest relative to the clustering around a random matter particle. In the velocity domain, an ``inner depletion radius", $\rid$, can be defined at the location of the maximum mass inflow rate. According to continuity, the maximum inflow radius marks the transition between an inner growing halo and the depleting environment, or the inner edge of the \emph{active} depletion region. 

These two radii are closely related, with $\rd$ slightly larger than $\rid$ by $10-20\%$ over the mass range we studied. Matter is drained from the environment outside $\rid$, leading to the formation of an accretion trough in the bias profile around $\rd$, which also corresponds to a relatively flat shoulder in the density profile beyond $\rd$. The bias profile and the velocity profile are also complementary for the practical identification of the depletion scale, with $\rd$ clear in the bias profile for low mass haloes but $\rid$ easier to identify in the velocity profile for cluster haloes. As the halo density profile typically has a logarithmic slope close to $-2$ around these scales, this $\rid$ can also be identified as the maximum infall velocity location.

We study how $\rd$ depends on multiple halo properties including halo mass, formation time, concentration, shape, spin and environment by binning the bias profiles according to one or two halo properties. $\rd$ in the binned profiles depends strongly on both mass and environment. The formation time and concentration dependence is also clear and non-redundant from the mass dependence, while the dependence on halo spin is the weakest. $\rd$ likely also depends on halo shape beside mass and formation time. 

Comparing $\rd$ to the splashback radius defined at the steepest slope location, we find $\rd$ is $\sim 2-3$ times the splashback radius. The ratio between the two increases with decreasing halo size and depends on halo properties in a way similar to the dependence of the NFW concentration on halo properties, with low mass and early forming haloes also having a larger ``outer concentration'' according to the depletion to splashback ratio. By contrast, the ratio between $\rd$ and the virial radius is $\sim 1-3$ and depends on halo properties in a very different way. The depletion to virial radius ratio is independent of mass ($\rd=2.5r_{\rm vir}$ at fixed mass) but sensitive to other halo parameters, and the trends are different from those for the depletion to splashback ratio. These results reflect that the different radii carry different information about the haloes. As $\rd$ is on average much farther out than the virial and splashback radius, it is subject to more influence from the environmental density.

Comparing $\rd$ to the turnaround radius, which can be interpreted as the outer depletion edge encompassing the infall region around haloes, we find $\rd$ approaches the turnaround in low mass haloes ($M_{\rm vir}\sim 10^{12}\mathrm{M}_\odot/h$ and below) that have stopped mass accretion on turnaround scales and reached their maximum turnaround radius. These haloes have well-formed bias troughs but are not surrounded by an infall region due to their completed mass accretion, leading to the agreement between $\rd$ and the turnaround radius. By contrast, massive haloes are surrounded by a clear infall region within the turnaround radius, while the bias trough is much shallower or absent, reflecting their younger age in the mass accretion process. At the high mass end, $\rd$ is thus located within the turnaround radius.

We have further studied the depletion radius in the phase space diagram of haloes, and find it naturally marks the transition between an inner halo structure and the outer environment according to the distribution of radial velocity at different radius. In particular, the inner depletion radius, $\rid$, can be interpreted as the radius enclosing a highly complete population of splashback orbits, or the radius of the splashback-bounded halo. Such high percentile splashback radius is not easy to robustly quantify from particle dynamics alone, while our definition via the maximum infall provides a natural alternative.

We find the average density enclosed in $\rd$ is approximately $40$ times the background density of the Universe, independent of the virial mass but dependent on other halo properties. Early forming, highly concentrated, spherical and high spin haloes tend to have a smaller $\rd$ at a fixed mass. Correspondingly, such haloes also tend to have a higher enclosed density, irrespective of mass. For haloes with $M_{\rm vir}>10^{12}\mathrm{M}_\odot/h$, the density enclosed in $\rid$ is also approximately a constant of $\sim 63$ times the background density.

As the depletion radii mark the outer edges of matter associated with the halo, they are expected to be applicable in the halo model description of the large-scale structure. We demonstrate this by comparing our new radii to the optimal halo exclusion radius found in the recent work of~\citet{Garcia2020}, who solved for the optimal halo exclusion radius by fitting a flexible halo model to the halo-matter correlation function. We found that the inner depletion radius, $\rid$, is in excellent agreement with their optimal exclusion radius, while $\rd$ is in agreement with their ``by-eye" halo boundary. As our radii are defined according to physical interpretation of the spatial and velocity distribution around haloes without tuning, it is really exciting to see that our data-driven exploration of the halo radius converges with their model-driven definition of the halo boundary, signalling the convergence towards the beauty and power of a more physical characterisation of structures in the universe. 

In this work, we focus on introducing these characterisations and studying their properties in the present day Universe, with some extra efforts in characterising $\rd$. In a follow-up work, we will study the evolution of the bias and velocity profiles to examine the depletion process explicitly, and to gain a dynamical understanding of these radii and their further connections, with more focuses on $\rid$.

\section*{Acknowledgements}
The authors are grateful to Yipeng Jing for providing access to the CosmicGrowth simulation, and to Hongyu Gao, Zhaozhou Li, Haojie Xu, Ji Yao, Jiaxin Wang, and Lindsay King for helpful discussions. This work is supported by NSFC (11973032, 11890691, 11621303), National Key Basic Research and Development Program of China (No.2018YFA0404504) and 111 project No. B20019. We gratefully acknowledge the support of the Key Laboratory for Particle Physics, Astrophysics and Cosmology, Ministry of Education. The computation of this work is partly done on the \textsc{Gravity} supercomputer at the Department of Astronomy, Shanghai Jiao Tong University.

\section*{Data availability}
The data underlying this article will be shared on reasonable request to the corresponding author.
%%%%%%%%%%%%%%%%%%%%%%%%%%%%%%%%%%%%%%%%%%%%%%%%%%



\bibliographystyle{mnras}
\bibliography{ref} % if your bibtex file is called example.bib

%%%%%%%%%%%%%%%%%%%%%%%%%%%%%%%%%%%%%%%%%%%%%%%%%%

%%%%%%%%%%%%%%%%% APPENDICES %%%%%%%%%%%%%%%%%%%%%

\appendix

\bsp	% typesetting comment
\label{lastpage}
\end{document}